\documentclass{appolb}
\usepackage[top=3cm, bottom=3cm, left=3cm, right=3cm]{geometry}
\usepackage{amsmath}
\usepackage{amsfonts}
\usepackage{amssymb}
\usepackage{graphicx}
\graphicspath{{FiguresPDF/}}
\usepackage[colorlinks,citecolor=blue,linktoc=all,linkcolor=cyan]{hyperref}
\usepackage{graphicx}
\usepackage{svg}
\usepackage{mathtools}
\usepackage{upgreek}
\usepackage{physics}
\usepackage{multirow}
\usepackage{placeins} 
\usepackage{pgfplots}
\usepackage{subfigure}
\usepackage{caption}
\usepackage{array}

\usepackage[T1]{fontenc}
\usepackage{dsfont}               
\usepackage{mathrsfs}             
\usepackage{slashed}              
\usepackage{amsbsy}
\usepackage{braket}
\usepackage{tikz}
\usepackage{bbold}

\renewcommand{\thetable}{\Roman{table}}

\providecommand{\keywords}[1]
{	
  \textbf{\textit{Keywords---}} #1
}

\usepackage[
    backend=biber,
    citestyle=numeric-comp,
    bibstyle=science,
    sorting=none,
    eprint=true,
    url=false,
    doi=true]{biblatex}
\DeclareFieldFormat{eprint:arxiv}{%
  Arxiv:\space
  \href{https://arxiv.org/abs/#1}{#1}%
  \iffieldundef{eprintclass}{}{ \mkbibbrackets{\thefield{eprintclass}}}%
}

\renewbibmacro*{doi+eprint+url}{%
  \printfield{doi}%
  \iffieldundef{doi}{%
    \iffieldundef{journaltitle}{%
      \setunit{\addspace}%
      \printfield[eprint:arxiv]{eprint}%
    }{}%
  }{}%
}

\DeclareSourcemap{
  \maps[datatype=bibtex]{
    \map{
      \step[fieldset=title, null]
    }
  }
}

\usepackage{xpatch}
\xpatchbibmacro{name:andothers}{ 
{\finalandcomma}%
}{%
\addspace%
}{}{}

\DeclareFieldFormat{doi}{%
  \mkbibacro{DOI}\addcolon\space
  \ifhyperref
    {\href{https://doi.org/#1}{\nolinkurl{#1}}}
    {\nolinkurl{#1}}}

\DeclareFieldFormat{labelnumberwidth}{\mkbibbrackets{#1}}

\AtEveryBibitem{\clearfield{month}}

\providecommand{\U}[1]{\protect\rule{.1in}{.1in}}

\pgfplotsset{compat=1.8}

\usepackage{authblk}          

\title{Two-exponential decay of Acridine Orange }

\author[1,2]{Francesco Giacosa\thanks{corresponding author: fgiacosa@ujk.edu.pl}}
\author[1]{Anna Kolbus}
\author[1]{Krzysztof Kyziol}
\author[1]{Magdalena Plodowska}
\author[1]{Milena Piotrowska}
\author[1]{Karol Szary}
\author[1]{Arthur Vereijken}

\affil[1]{Jan Kochanowski University, ul. \.Zeromskiego 5, 25-369 Kielce, Poland}
\affil[2]{Goethe University, Theodor-W.-Adorno-Platz 1, 60323, Frankfurt am Main, Germany}

\date{\today}

\begin{document}

\sloppy 
\maketitle

\begin{abstract}
    In this work, we experimentally study the fluorescence decay of Acridine Orange at late times, in order to test whether a late-time power-law behaviour emerges — a feature expected to be very small but consistent with quantum mechanical and quantum field theoretical predictions. Using two distinct photon detectors, we find that the data are well described by a sum of two exponential functions with lifetimes $\tau_1 = 1.7331 \pm 0.001$ ns and $\tau_2 = 5.948 \pm 0.012$ ns, in agreement with values reported in the literature. While no deviation from the exponential decay law is observed, this study serves as a reliable test for the experimental setup and enables a precise determination of the sample lifetimes.
\end{abstract}

\bigskip

\textbf{Introduction: }
The standard radioactive decay law describes the change in number of elements in a sample as $N(t)=N_0 \,e^{-t/\tau},$ an exponential function. However, this successful phenomenological formulation is not strictly reproduced within the framework of Quantum Mechanics (QM) and Quantum Field Theory (QFT) according to which $N(t) = N_0 P(t)$, where $P(t)$ is the survival probability as a function of time $t$ of a single unstable quantum state. The quantity $P(t)$ emerges as the modulus squared of the Fourier transform of the energy distribution $\rho (E)$ of the unstable quantum state/particle, e.g. \cite{LFonda_1978}:
\begin{equation}
    \label{eq:survival-probability}
    P(t)=\left| \int_{E_{th}}^{\infty} \mathrm{d}E \, \rho (E) e^{-i\frac{E}{\hbar} t }\right| ^2 \; \text{ ,} 
\end{equation}
 where $E_{th}$ is the lowest admissible energy. Only for $E_{th} \rightarrow - \infty$ and $\rho(E)$ being a Breit-Wigner distribution, the decay is purely exponential. 
 If, instead, the expectation value of the Hamiltonian operator $H$ is finite, then $P'(0)=0$. If also the $H^2$-expectation is such, $P(t)$ can be approximated at short times as $P(t)\simeq 1-t^2/t_Z^2$ with $\tau_Z = \hbar / \sigma_E$, where $\sigma _E$ is the standard deviation of the energy distribution. This deviation from the exponential decay law at short times is linked to the quantum Zeno effect (QZE), i.e., the freezing in its initial state if probed sufficiently often \cite{Misra:1976by, 1998PhRvA..57.1509S,,Facchi:2008nrb, 2002OptCo.211..235B,Kofman:2000gle}. On the other hand, at large times the survival probability $P(t)$ is well described by a power law, $P(t) \sim t^{-(\beta -1)}$, leading to the decay rate intensity $I(t) \sim t^{-\beta }$. This is a direct consequence of the presence of the energy threshold $E_{th}$ \cite{khalfin}. 
 Both mentioned effects are predicted in the framework of QM \cite{LFonda_1978,dicus2002,Urbanowski1994} and also quantum field theory (QFT) \cite{Giacosa:2011xa,Giacosa:2010br,Giacosa:2021hgl,Facchi:1999ik}  and emerge from their very first principles. However, the deviations are typically very small for elementary systems/transitions, as the `parade' example of the $2P$-$1S$ transition of the $H$-atom shows: short-time deviations take place at time $10^{-8}\tau$ and late-time deviations for $\sim 125 \tau$ ($\tau = 1.595$ ns), thus very early and very late, rendering an observation extremely challenging \cite{Facchi:1998abc,Giacosa:2024yhn}.

 At short times, the experimental verification of deviations from the exponential law was measured in the study of quantum tunneling of sodium atoms within an optical potential \cite{1997Natur.387..575W,2001PhRvL..87d0402F}. (For indirect evidence, see the photonic-waveguide array experiment reported in Ref. \cite{crespi}; strongly decay systems also display deviations from the Breit-Wigner function \cite{ALEPH:2005qgp,Giacosa:2010br}, 
 in turn implying a nonexponential behaviour which, however, cannot be directly measured due to the extremely short times ($\sim 10^{-23}$ s) involved.)

For long times, the well-known work of Ref. \cite{rothe} reports a power law intensity $I(t)$ for various chemical compounds decaying via fluorescence  \cite{lakowicz}. Very recently, a power law was confirmed by our group for the case of erythrosine B upon using two distinct photodetectors \cite{Giacosa:2025knr}. In both cases, dissolved fluorophores were employed; such systems, even if not as simple as `more elementary quantum states'  because of inhomogeneous broadening of the emission spectrum resulting from interaction of the compound with the solvent, are thought to still be described by the inherent quantum Eq. (\ref{eq:survival-probability}).

In this work, we report on the late-time study of a different substance, acridine orange, measured using two photon detectors acting in different ranges. As we shall show, the decay rate can be well described by the sum of two exponential functions, in agreement with a mixture of two quantum states with two distinct lifetimes. The rather precise determination of lifetime(s) is in agreement with values reported in the literature. Moreover, even if no `quantum' late-time memory effect can be seen,  this study offers a valuable test of our experimental approach aimed at investigating the late-time decay law.  

\textbf{Basic features of  fluorescence.} Excited states of fluorophores provide typical examples of unstable quantum states. These molecules may absorb a photon, that implies an excitation of an electron to higher energy levels denoted by $S_1,S_2,S_3$, ... . Then, this electron undergoes a series of rapid transitions via a non-radiative process referred to as vibrational relaxation. After a short time ($\sim 10^{-12} \; \mathrm{s}$), the electron is in the lowest vibrational level of the excited electronic state. Subsequently, it may decay into lower electronic states, usually via internal conversion (IC) -- which is also a non-radiative process -- or via the radiative fluorescence process, which we detect. This transition usually occurs from the lowest vibrational level of the first excited electronic state ($S_1$) to one of many vibrational levels of the electronic ground state ($S_0$). According to the Franck-Condon rule, the most probable transitions are those for which the initial and final electronic wave functions overlap the most. Typically, the dynamics of the fluorescence takes place in a timescale of the order of nanoseconds.     

\textbf{Description of the device. }
In the performed experiment, the Time-Correlated Single Photon Counting (TCSPC) setup available at the Institute of Biology, Faculty of Natural Sciences, Jan Kochanowski University in Kielce (Poland) was used. It consists of PicoQuant Laser Combining Unit (LCU), Nikon Eclipse Ti-E Inverted Confocal Microscope, detection system, and PicoQuant PicoHarp 300 TCSPC module. 
The laser combining unit is composed of two picosecond laser diodes -- PicoQuant LDH-D-C-440 and PicoQuant LDH-D-C-485 emitting at, respectively, 438 nm and 485 nm with spectral bandwidth between 2 and 8 nm. However, only the latter one was actually used during the measurements. Laser diodes serve as pulsed excitation sources. They were operating at 10 MHz frequency (100 ns interval between pulses) and the width of the pulse at half maximum (FWHM) is less than 120 ps. 

The detection setup is composed of two identical PicoQuant PMA Hybrid 40 detectors. They are separated with the dichroic mirror and every detector is coupled to an individual bandpass filter -- FF01-520/35 and ET600/50. Thus, two independent detection channels can be distinguished that focus at different parts of the spectrum -- Channel 1 detects photons of wavelength ranging from 485 to 555 nm and Channel 2 photons from 550 to 650 nm. \
Fluorescent samples are placed on the stage of the Nikon Eclipse Ti-E Inverted Confocal Microscope. Thus, the microscope plays an important role in the formation of the excitation and emission beams and their guidance. In our case, it is equipped with one optical fibre transferring the laser pulse from the LCU to the device and another one extracting the fluorescence signal towards the detection system. The last important piece is the PicoHarp 300 TCSPC module which is responsible for timing events, Analog-to-Digital conversion of the signal, and, finally, assigning particular events to appropriate time bins of the histogram.

\textbf{Description of the experiment.}
The choice of acridine orange was motivated by the shape of the absorption spectrum that matches the available excitation wavelengths of the LCU, reasonable quantum yields, and its common use in fluorescence studies, offering a valuable test of our device. 

As a first step, the fluorophore was dissolved in water reaching a concentration equal to $10^{-5} \; \mathrm{mol/dm^3}$. 
After transferring the sample to the confocal microscope, one real-time test measurement was performed to locate the maximum signal-to-noise ratio. This initial procedure was followed by three 10-minute measurements, each of them made on a different part of the sample. The geometrical conditions remained similar throughout the process. Thus, the obtained fluorescence intensity decay curves could be combined into one final data set for an individual detection channel. 

\textbf{Results for Acridine Orange.} In the process of analysis, the functions describing the decay rate were obtained by a $\chi^2$-approach. The standard deviation was estimated with Poisson statistics. Thus, for a given rate $I(t)$, the $\chi ^2$ reads:
\begin{equation}
    \label{eq:chi-squared}
    \chi ^2=\sum_{n=1}^{N_t} \frac{\left[ I_n- I(t_n)\right] ^2}{I(t_n)} \; ,
\end{equation}
where $N_t$ is the total number of data points. A single exponential function was tested, but it cannot describe the data. Hence, we employed two models -- a sum of two exponential functions and a nonexponential late-time power law. In the first model, it is assumed that there are two types of initial excited states which decay exponentially, characterised by different lifetimes ($\tau_1$ and $\tau _2$). The second encodes a late-time QM-inspired power law. 

\begin{table}[!htbp]
    \centering
    \caption[Model functions employed in the analysis.]{Model functions employed in the analysis; $b$ is the background. The time $t_0 = 2.24$ ns corresponds (for both channels) to the maximum of the intensity (and is not a fit parameter).}
    {\renewcommand{\arraystretch}{2}
    \begin{tabular}{|c|c|c|}
         \hline
         Model & Fluorescence intensity $I(t)$ & Fit parameters \\ \hline
         Two-exponential & $I(t)=C_1\exp \left( -\frac{t-t_0}{\tau_1} \right) + C_2\exp \left( -\frac{t-t_0}{\tau_2} \right) + b$ & $\chi ^2 (C_1,\tau _1, C_2,\tau _2, b)$ \\ \hline
         Nonexponential & $I(t)=C \exp \left( -\frac{t-t_0}{\tau} \right) + C_p \, (t-t_0)^{-\beta} +b$ & $\chi ^2 (C_0,\tau, C_p,\beta, b)$ \\ \hline
    \end{tabular}}
    \label{tab:models}
\end{table}
For completeness, we recall that the  fluorescence intensity $I(t)$ is related to the survival probability $P(t)$ of Eq. (1) as 
\begin{equation}
    I(t)=-\frac{\mathrm{d}N}{\mathrm{d}t} = -N_0 \, P'(t)\; .
    \label{eq:fluorescence_intensity-definition}
\end{equation}
.
\begin{table}[!htbp]
    \centering
    \captionsetup{justification=centering}
    \caption{Fitting results for Acridine Orange measurements.}
    \label{tab:data-analysis}
    \renewcommand{\arraystretch}{1.4}
    \begin{tabular}{|c|c|c|c|c|c|c|}
    \hline
    \multicolumn{7}{|c|}{Fitting Range: 3.200 -- 96.968 ns} \\ \hline
    \multicolumn{7}{|c|}{Two-exponential model} \\ \hline
    Channel & $\chi_{\nu}^2$ &$C_1$ & $\tau_1$ [ns] &  $C_2$ &$\tau_2$ [ns] & $b$ \\ \hline
    1 & 1.0471 & 278 967 & 1.7333 & 6 371.3 & 5.9459 & 21.99 \\ \hline
    2 & 1.0247 & 150 898 & 1.7326 & 8 983.5 & 5.9493 & 42.07 \\ \hline
    \multicolumn{7}{|c|}{Nonexponential model} \\ \hline
    Channel & $\chi_{\nu}^2$ &$C$ & $\tau$ [ns] &  $C_p$ &$\beta$ [ns] & $b$ \\ \hline
    1 & 11.0881 & 220 113 & 1.9360 & 39 458 & 1.7386 & -1.404 \\ \hline
    2 & 14.7270 & 93 100 & 2.2649 & 40 747 & 2.421 & 2.421 \\ \hline
    \end{tabular}
\end{table}

\begin{figure}[!htbp]
    \centering
    \subfigure{
        \includegraphics[scale=1.25]{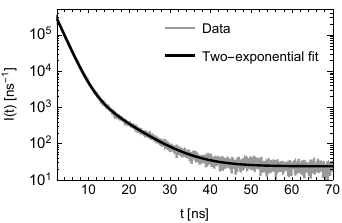}}
    \hfill
     \subfigure{
        \includegraphics[scale=1.25]{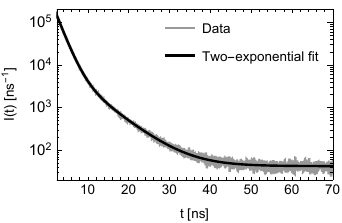}}
    \caption{Fluorescence intensity for both photon detectors (channel 1: left, channel 2: right) - comparison between data and two-exponential fitting function.}
\end{figure}

\FloatBarrier

\textbf{Discussions.}
The values of the fit parameters that minimise $\chi^2$ were found numerically\footnote{Since the excitation is not instantaneous, we start the fit when the early-time effects are not relevant. In the case of the nonexponential model, this is also required by the fact that the power function is not a valid approximation in the early-time domain.}, see the results in Tab. \ref{tab:data-analysis}.  The following comments are in order:
\begin{itemize}
    \item For acridine orange the two-exponential model works well, see the direct comparison of the data and the fit in Fig. 1. The two-exponential behaviour of the decay was previously observed in aqueous solutions of Acridine Orange in sodium dodecyl sulfate (SDS) as a consequence of formation of the micellar structures due to interaction with the detergent -- see Refs. \cite{AO1983,AO1988}. The values of both lifetimes are also in the highlighted range (1-2 ns for the main and >3 ns for the other lifetime). This fact suggests that we observe similar aggregates. Interestingly, the deviations from single exponential decay law were also observed earlier in the case of proflavine, which is an acridine's derivative, bound to DNA \cite{Georghiou} and for acridine orange itself \cite{KUBOTA1977279}. However, in the latter case this effect was caused mainly by the response of the instrument, and the authors observed a single-lifetime exponential decay after deconvolution of the signal.

    \item Following a conservative estimate we quote the following lifetime determinations (inverse-variance weighted mean):
    \begin{equation}
       \tau_1 = 1.7331 \pm 0.001~\text{ns , }  \tau_2 = 5.948 \pm 0.012~\text{ns .} 
    \end{equation}
   The lifetimes determined by the different detectors are consistent with each other: for Channel 1, the standard deviations are $0.0012~\text{ns}$ for $\tau_1$ and $0.0196~\text{ns}$ for $\tau_2$; for Channel 2 they are $0.0021~\text{ns}$ and $0.0156~\text{ns}$, respectively. The errors of the fit parameters were taken as the diagonal elements of the covariance matrix (the inverse of the Hesse matrix) \cite{cowan}. As summarized in Table~\ref{tab:data-analysis}, the absolute channel-to-channel differences are $|\Delta\tau_1|=0.0007~\text{ns}$ and $|\Delta\tau_2|=0.0034~\text{ns}$, both comfortably below the quoted conservative error bars. A noticeable change is observed only in the $C_1/C_2$ ratio, attributable to the spectral shape and to the position of the spectral window associated with each detection channel.

    \item The nonexponential model does not offer an acceptable data description, see Tab. \ref{tab:data-analysis}. Thus, no sign of a QM power-law is seen in the present data for acridine orange, even if we follow the sample up to more than $10 \; \tau_1$. 
    (Note, the eventual appearance of a power law after about 10 lifetimes \cite{rothe,Giacosa:2025knr} implies the existence of an effective threshold $E_th$ quite close to the resonance peak \cite{Giacosa:2025knr}. Yet, the exact nature and emergence of this threshold is not fully understood, so at present one cannot predict the turnover time but can only measure it.)  

    \item  In Ref. \cite{Giacosa:2025knr}, using the same setup we could measure a power-law for erythrosine B. The comparison of that system with the one of this paper allows us to understand which features favour the detection of the late-time power-law: (i) erythrosine B main lifetime is shorter ($\tau_1 = 0.45$ ns), thus we could measure the decay curve $I(t)$ for a larger $t/\tau_1$ fraction. In general, a dye lifetime less than 1 ns is definitely preferable for our purposes. (ii) Erythrosine B is dominated by a single lifetime plus a late-time power law. Acridine Orange exhibits two lifetimes originating from a two-state admixture, the longer one dominating at times where a late-time power-law behaviour would possibly emerge.

    \item For erythrosine B, a two-exponential fit was tested, but it is definitely worse than the power-law one. Moreover, the two detectors would measure incompatible values for $\tau_2$ \cite{Giacosa:2025knr}, which makes a two-state interpretation untenable. Indeed, also the power-law coefficients $\beta$ measured by both channels differ, but this is consistent with basic QM and QFT expectations \cite{Giacosa:2021hgl,Giacosa:2011xa}.  
\end{itemize}

\textbf{Conclusions.} In this paper, we have presented decay–intensity measurements of acridine orange, whose fluorescence is well described by the sum of two exponential components, consistently confirmed by both photon detectors. Although no late-time power-law behaviour has been observed, this study provides a valuable validation of our experimental setup for late-time decay measurements, as discussed in Ref.~\cite{Giacosa:2025knr}, and defines the experimental conditions necessary for future searches of possible quantum-mechanical (and quantum-field-theoretical) deviations from the exponential decay law.

\bigskip 

\textbf{Acknowledgments}
This work was supported by the Polish Minister of Science under the ‘Regional Excellence Initiative’ program (project RID/SP/0015/2024/01, sub-projects RID/2024/LIDER/08 and RID/2025/LIDER/02).

\FloatBarrier

\printbibliography

@article{Facchi:1998abc,
    author = "Facchi, P. and Pascazio, S.",
    title = "{Temporal behavior and quantum zeno time of an excited state of the hydrogen atom}",
    eprint = "quant-ph/9905017",
    archivePrefix = "arXiv",
    doi = "10.1016/S0375-9601(98)00144-3",
    journal = "Phys. Lett. A",
    volume = "241",
    pages = "139--144",
    year = "1998"
}

@article{LFonda_1978,
    doi = {10.1088/0034-4885/41/4/003},
    url = {https://dx.doi.org/10.1088/0034-4885/41/4/003},
    year = {1978},
    month = {apr},
    publisher = {},
    volume = {41},
    number = {4},
    pages = {587},
    author = {L Fonda and  G C Ghirardi and  A Rimini},
    title = {Decay theory of unstable quantum systems},
    journal = {Rep. Prog. Phys.}
}

@article{Giacosa:2011xa,
    author = "Giacosa, Francesco",
    title = "{Non-exponential decay in quantum field theory and in quantum mechanics: the case of two (or more) decay channels}",
    eprint = "1110.5923",
    archivePrefix = "arXiv",
    primaryClass = "nucl-th",
    doi = "10.1007/s10701-012-9667-3",
    journal = "Found. Phys.",
    volume = "42",
    pages = "1262--1299",
    year = "2012"
}

@article{Giacosa:2021hgl,
    author = "Giacosa, Francesco",
    title = "{Multichannel decay law}",
    eprint = "2108.07838",
    archivePrefix = "arXiv",
    primaryClass = "quant-ph",
    doi = "10.1016/j.physletb.2022.137200",
    journal = "Phys. Lett. B",
    volume = "831",
    pages = "137200",
    year = "2022"
}

@article{Kofman:2000gle,
    author = "Kofman, A. G. and Kurizki, G.",
    title = "{Acceleration of quantum decay processes by frequent observations}",
    doi = "10.1038/35014537",
    journal = "Nature",
    volume = "405",
    number = "6786",
    pages = "546--550",
    year = "2000"
}

@article{Misra:1976by,
    author = "Misra, B. and Sudarshan, E. C. G.",
    title = "{The Zeno's Paradox in Quantum Theory}",
    reportNumber = "ORO-3992-271",
    doi = "10.1063/1.523304",
    journal = "J. Math. Phys.",
    volume = "18",
    pages = "756",
    year = "1977"
}

@ARTICLE{rothe,
       author = {{Rothe}, C. and {Hintschich}, S.~I. and {Monkman}, A.~P.},
        title = "{Violation of the Exponential-Decay Law at Long Times}",
      journal = {Phys. Rev. Lett.},
         year = 2006,
        month = apr,
       volume = {96},
       number = {16},
          eid = {163601},
        pages = {163601},
          doi = {10.1103/PhysRevLett.96.163601},
       adsurl = {https://ui.adsabs.harvard.edu/abs/2006PhRvL..96p3601R}
}

@ARTICLE{1997Natur.387..575W,
       author = {{Wilkinson}, Steven R. and {Bharucha}, Cyrus F. and {Fischer}, Martin C. and {Madison}, Kirk W. and {Morrow}, Patrick R. and {Niu}, Qian and {Sundaram}, Bala and {Raizen}, Mark G.},
        title = "{Experimental evidence for non-exponential decay in quantum tunnelling}",
      journal = {Nature},
         year = 1997,
        month = jun,
       volume = {387},
       number = {6633},
        pages = {575-577},
          doi = {10.1038/42418},
       adsurl = {https://ui.adsabs.harvard.edu/abs/1997Natur.387..575W}
}

@article{Facchi:2008nrb,
    author = "Facchi, P. and Pascazio, S.",
    title = "{Quantum Zeno dynamics: mathematical and physical aspects}",
    doi = "10.1088/1751-8113/41/49/493001",
    journal = "J. Phys. A",
    volume = "41",
    number = "49",
    pages = "493001",
    year = "2008"
}

@ARTICLE{1998PhRvA..57.1509S,
       author = {{Schulman}, L.~S.},
        title = "{Continuous and pulsed observations in the quantum Zeno effect}",
      journal = {Phys. Rev. A},
         year = 1998,
        month = mar,
       volume = {57},
       number = {3},
        pages = {1509-1515},
          doi = {10.1103/PhysRevA.57.1509},
       adsurl = {https://ui.adsabs.harvard.edu/abs/1998PhRvA..57.1509S}
}

@ARTICLE{2002OptCo.211..235B,
       author = {{Balzer}, Chr. and {Hannemann}, Th. and {Rei{\ss}}, D. and {Wunderlich}, Chr. and {Neuhauser}, W. and {Toschek}, P.~E.},
        title = "{A relaxationless demonstration of the Quantum Zeno paradox on an individual atom}",
      journal = {Opt. Commun.},
         year = 2002,
        month = oct,
       volume = {211},
       number = {1-6},
        pages = {235-241},
          doi = {10.1016/S0030-4018(02)01859-X},
archivePrefix = {arXiv},
       eprint = {quant-ph/0406027},
 primaryClass = {quant-ph},
       adsurl = {https://ui.adsabs.harvard.edu/abs/2002OptCo.211..235B}
}

@ARTICLE{2001PhRvL..87d0402F,
       author = {{Fischer}, M.~C. and {Guti{\'e}rrez-Medina}, B. and {Raizen}, M.~G.},
        title = "{Observation of the Quantum Zeno and Anti-Zeno Effects in an Unstable System}",
      journal = {Phys. Rev. Lett.},
         year = 2001,
        month = jul,
       volume = {87},
       number = {4},
          eid = {040402},
        pages = {040402},
          doi = {10.1103/PhysRevLett.87.040402},
archivePrefix = {arXiv},
       eprint = {quant-ph/0104035},
 primaryClass = {quant-ph},
       adsurl = {https://ui.adsabs.harvard.edu/abs/2001PhRvL..87d0402F}
}

@ARTICLE{crespi,
       author = {{Crespi}, Andrea and {Pepe}, Francesco V. and {Facchi}, Paolo and {Sciarrino}, Fabio and {Mataloni}, Paolo and {Nakazato}, Hiromichi and {Pascazio}, Saverio and {Osellame}, Roberto},
        title = "{Experimental Investigation of Quantum Decay at Short, Intermediate, and Long Times via Integrated Photonics}",
      journal = {Phys. Rev. Lett.},
         year = 2019,
        month = apr,
       volume = {122},
       number = {13},
          eid = {130401},
        pages = {130401},
          doi = {10.1103/PhysRevLett.122.130401},
archivePrefix = {arXiv},
       eprint = {1903.05378},
 primaryClass = {quant-ph},
       adsurl = {https://ui.adsabs.harvard.edu/abs/2019PhRvL.122m0401C}
}

@article{Facchi:1999ik,
    author = "Facchi, P. and Pascazio, S.",
    title = "{Van Hove's '$\lambda^2t$' limit in nonrelativistic and relativistic field theoretical models}",
    eprint = "quant-ph/9910111",
    archivePrefix = "arXiv",
    reportNumber = "BA-TH-99-366",
    doi = "10.1016/S0960-0779(01)00090-X",
    journal = "Chaos Solitons Fractals",
    volume = "12",
    pages = "2777",
    year = "2001"
}

@article{Giacosa:2010br,
    author = "Giacosa, Francesco and Pagliara, Giuseppe",
    title = "{Deviation from the exponential decay law in relativistic quantum field theory: the example of strongly decaying particles}",
    eprint = "1005.4817",
    archivePrefix = "arXiv",
    primaryClass = "hep-ph",
    doi = "10.1142/S021773231103670X",
    journal = "Mod. Phys. Lett. A",
    volume = "26",
    pages = "2247--2259",
    year = "2011"
}

@article{Urbanowski1994,
  title = {Early-time properties of quantum evolution},
  author = {Urbanowski, K.},
  journal = {Phys. Rev. A},
  volume = {50},
  issue = {4},
  pages = {2847--2853},
  numpages = {0},
  year = {1994},
  month = {Oct},
  publisher = {APS},
  doi = {10.1103/PhysRevA.50.2847},
  url = {https://link.aps.org/doi/10.1103/PhysRevA.50.2847}
}

@article{Giacosa:2025knr,
    author = "Giacosa, Francesco and Kolbus, Anna and Kyziol, Krzysztof and Plodowska, Magdalena and Piotrowska, Milena and Szary, Karol and Vereijken, Arthur",
    title = "{Quantum Late-Time Decay and Channel Dependence}",
    eprint = "2509.17163",
    archivePrefix = "arXiv",
    primaryClass = "quant-ph",
    month = "9",
    year = "2025"
}

@book{lakowicz,
author = {Lakowicz, J.},
year = {2006},
month = {01},
pages = {},
title = {Principles of Fluorescence Spectroscopy, 3rd Edition},
volume = {1},
isbn = {978-0-387-31278-1},
publisher = {Springer},
doi = {10.1007/978-0-387-46312-4}
}

@article{Giacosa:2024yhn,
    author = "Giacosa, F. and Kyzio{\l}, K.",
    title = "{Nonexponential Decay Law of the 2P{\textendash}1S~Transition of the H Atom}",
    eprint = "2408.06905",
    archivePrefix = "arXiv",
    primaryClass = "quant-ph",
    doi = "10.12693/aphyspola.146.704",
    journal = "Acta Phys. Polon. A",
    volume = "146",
    number = "5",
    pages = "704--708",
    year = "2024",

}

@article{khalfin,
  author       = {Khalfin, L. A.},
  title        = {ON THE THEORY OF THE DECAY AT QUASI-STATIONARY STATE},
  journal      = {Doklady Akad. Nauk S.S.S.R.},
  volume       = {Vol: 115},
  place        = {Country unknown/Code not available},
  year         = {1957},
  month        = {07}
}

@book{cowan,
author = {Cowan, Glen},
publisher = {Oxford University Press},
year = {2023},
month = {10},
pages = {},
title = {Statistical Data Analysis},
isbn = "978-0-19-850156-5",
doi = {10.1093/oso/9780198501565.001.0001}
}

@article{AO1983,
author = {Ban, Taketora and Kasatani, Kazuo and Kawasaki, Masahiro and Sato, Hiroyasu},
title = {FLUORESCENCE DECAY OF THE ACRIDINE ORANGE-SODIUM DODECYL SULFATE SYSTEM: FORMATION OF DYE-RICH INDUCED MICELLES IN THE PREMICELLAR REGION},
journal = {Photochem. Photobiol.},
volume = {37},
number = {2},
pages = {131-139},
doi = {https://doi.org/10.1111/j.1751-1097.1983.tb04448.x},
url = {https://onlinelibrary.wiley.com/doi/abs/10.1111/j.1751-1097.1983.tb04448.x},
eprint = {https://onlinelibrary.wiley.com/doi/pdf/10.1111/j.1751-1097.1983.tb04448.x},
year = {1983}
}

@article{AO1988,
author = {Miyoshi, N. and Hara, K. and Yokoyama, I. and Tomita, G. and Fukuda, M.},
title = {Fluorescence lifetime of acridine orange in sodium dodecyl sulfate premicellar solutions},
journal = {Photochem Photobiol.},
year = {1988},
month = {05},
doi = {10.1111/j.1751-1097.1988.tb02765.x},
pages = {685-8}
}

@article{dicus2002,
  author       = {Dicus, Duane A. and Repko, Wayne W. and Schwitters, Roy F. and Tinsley, Todd M.},
  title        = {Time development of a quasistationary state},
  journal      = {Phys. Rev. A},
  volume       = {65},
  number       = {3},
  pages        = {032116},
  year         = {2002},
  doi          = {10.1103/PhysRevA.65.032116},
  eprint       = {quant-ph/0109053},
  archivePrefix= {arXiv},
  primaryClass = {quant-ph}
}

@article{ALEPH:2005qgp,
    author = "Schael, S. and others",
    collaboration = "ALEPH",
    title = "{Branching ratios and spectral functions of tau decays: Final ALEPH measurements and physics implications}",
    eprint = "hep-ex/0506072",
    archivePrefix = "arXiv",
    doi = "10.1016/j.physrep.2005.06.007",
    journal = "Phys. Rept.",
    volume = "421",
    pages = "191--284",
    year = "2005"
}

@article{KUBOTA1977279,
title = {Fluorescence decay and quantum yield characteristics of acridine orange and proflavine bound to DNA},
journal = {Biophysical Chemistry},
volume = {6},
number = {3},
pages = {279-289},
year = {1977},
issn = {0301-4622},
doi = {10.1016/0301-4622(77)85009-6},
url = {https://www.sciencedirect.com/science/article/pii/0301462277850096},
author = {Yukio Kubota and Robert F. Steiner}
}

@article{Georghiou,
author = {Georghiou, S.},
title = {ON THE NATURE OF INTERACTION BETWEEN PROFLAVINE and DNA},
journal = {Photochemistry and Photobiology},
volume = {22},
number = {3-4},
pages = {103-109},
doi = {10.1111/j.1751-1097.1975.tb08820.x},
url = {https://onlinelibrary.wiley.com/doi/abs/10.1111/j.1751-1097.1975.tb08820.x},
eprint = {https://onlinelibrary.wiley.com/doi/pdf/10.1111/j.1751-1097.1975.tb08820.x},
year = {1975}
}

\end{document}